\documentclass[aps,prd,twocolumn,showpacs,superscriptaddress,groupedaddress,eqsecnum,nofootinbib,floatfix]{revtex4-2}

\usepackage{amsmath}
\usepackage{amsfonts}
\usepackage{amssymb}	
\usepackage{graphicx}
\usepackage{bm}
\usepackage{color}
\usepackage{commath}

\usepackage{graphicx}
\usepackage{dcolumn}

\usepackage{color}
\usepackage{xcolor}
\definecolor{linkcolor}{rgb}{0.0,0.3,0.5}
\usepackage[unicode,colorlinks=true,citecolor=linkcolor,linkcolor=linkcolor,urlcolor=linkcolor]{hyperref}

\usepackage{cleveref}
\allowdisplaybreaks
\graphicspath{{../figures/}}

\begin{document}

\title{Gravitational scattering upto third post-Newtonian approximation for conservative dynamics: Scalar-Tensor theories}

\author{Tamanna \surname{Jain}$^{1,2}$}
\email{tj317@cam.ac.uk}
\affiliation{$^{1}$Department of Applied Mathematics and Theoretical Physics,University of Cambridge,Wilberforce Road CB3 0WA Cambridge, United Kingdom}%
\affiliation{$^{2}$Institut des Hautes Etudes Scientifiques, 91440 Bures-sur-Yvette, France}%
\date{\today}

\begin{abstract}

We compute the scattering angle $\chi$ for hyperboliclike encounters in massless Scalar-Tensor (ST) theories up to third post-Newtonian (PN) order for the conservative part of the dynamics.
To calculate the gauge-invariant scattering angle as a function of energy and orbital angular momentum, we use the approach of Effective-One-Body formalism as introduced in [Phys.Rev.D 96 (2017) 6, 064021].  We then compute the nonlocal-in-time contribution to the scattering angle by using the strategy of order-reduction of nonlocal dynamics introduced for small-eccentricity orbits. 
\end{abstract}

\maketitle
\section{Introduction}
The observation of gravitational wave signals, with the first observation by the LIGO-Virgo collaboration in 2015 \cite{LIGOScientific:2016aoc}, opened a new era to probe the dynamics of the strong-field gravity regime. The third generation of detectors \cite{Reitze:2019iox,Punturo:2010zz,Amaro-Seoane:2018gbb}, along with the next generation of telescopes, such as the Einstein telescope \cite{Maggiore:2019uih} and the Cosmic Explorer \cite{Evans:2021gyd} will be crucial for probing the strong-field dynamics of gravity by constraining the parameters of the alternative theories of gravity.

Amongst the theories alternative to Einstein's General Relativity (GR), the simplest theory is the addition of the massless scalar field to GR known as the scalar-tensor (ST) theory. The ST theories have been extensively studied and tested~\cite{Damour:1992we,Damour:1993hw,Damour:1995kt,Freire:2012mg,Khalil:2022sii,Gautam:2022cpb}. Besides arising naturally in the UV complete theories of gravity, the addition of the scalar field is also equivalent to $f(R)$-theories of gravity \cite{DeFelice:2010aj}. The two-body problem for ST theories have been extensively studied within the post-Newtonian (PN) approximation for both the dynamics and waveform generation in Refs.~\cite{Lang:2013fna,Lang:2014osa,Mirshekari:2013vb,Bernard:2019yfz,Bernard:2018hta,Bernard:2018ivi,Bernard:2022noq,Schon:2021pcv}. 

The detection of the gravitational wave signals relies on a large bank of (semi-) analytical accurate waveform templates to match filter against the data observed in the detectors. Therefore, the two-body PN dynamics in ST theories have been mapped within the effective-one-body (EOB) formalism~\cite{Julie:2017pkb,Julie:2017ucp,Jain:2022nxs,Jain:2023fvt,Julie:2022qux} to incorporate the corrections due to massless scalar-tensor theories in the EOB approach based waveform models \cite{Gamba:2021ydi,Gamba:2022mgx,Ossokine:2020kjp}. These results were obtained for the elliptic motions of the compact binaries. 

The EOB description of the unbound, scattering states of the binary systems was introduced in Ref.~\cite{Damour:2016gwp}. Recently, the approach was used to compute the scattering angle within the PN approximation in GR \cite{Bini:2017wfr}. The main aim of this paper is to compute the scattering angle in ST theories upto 3PN order using the EOB Hamiltonian in ST theories. 

The paper is organised as follows. In Sec.~\ref{Sec-reminder}, 
we give a brief reminder of ST theories and the EOB formalism in ST theories. 
Then, in Sec.~\ref{Sec-locscat} we derive the scattering angle for local part of the dynamics in ST theories at the 3PN order and derive the scattering angle for nonlocal part of the dynamics at 3PN order using order-reduction approach in Sec.~\ref{Sec-nonlocscat}. Finally, in Sec.~\ref{Sec-sum3PN} we sum the local and nonlocal contribution at 3PN in large-$j$ expansion.

\section{Brief Scalar-Tensor Theory  and EOB Reminder}
\label{Sec-reminder}
We consider mono-scalar massless ST theories described by the minimal coupling of the scalar field to the metric in the Einstein Frame, and its action reads
\begin{align}
S&=\frac{c^4}{16 \pi G}\int d^4x \sqrt{-g}(R-2g^{\mu \nu} \partial_\mu \varphi \partial_\nu \varphi)\nonumber\\
  &\qquad\qquad\qquad\qquad+S_m[\Psi, {\mathcal{A}(\varphi)}^2g_{\mu\nu}]~,
\end{align}
where $g_{\mu\nu}$ is the Einstein metric, $R$ is the Ricci scalar, $\varphi$ is the scalar field, $\Psi$ collectively denotes the matter fields, $g \equiv \det(g_{\mu\nu})$ and $G$ is the bare Newton's constant.
Here, we adopt the conventions and notations of Damour, Jaranowski and Sch\"{a}fer (DJS, hereafter)~\cite{Damour:1992we,Damour:1995kt}.

In Einstein Frame, the field equations for ST theories are derived in \cite{Damour:1992we}. The coupling of the scalar field with the matter fields gives rise to the dynamics of the scalar field where the coupling between the matter and the scalar field is measured by the parameter
\begin{equation}
\alpha(\varphi)=\frac{\partial \ln \mathcal{A}}{\partial \varphi} \ ,
 \end{equation}
in the equations of motion. The scalar field is non-minimally coupled to the metric in Jordan Frame (physical frame) 
\begin{equation}
\tilde{g}_{\mu\nu}={\mathcal{A}(\varphi)}^2 g_{\mu\nu} \ ,
\end{equation}
where $\tilde{g}_{{\mu\nu}}$ is the metric in Jordan frame, and hence the Einstein-Frame mass is defined as $m(\varphi)={\mathcal{A}(\varphi)}^2\tilde{m}(\varphi)$, where $\tilde{m}(\varphi)$ is the scalar-field dependent mass. The Jordan-Frame parameters of Ref.~\cite{Mirshekari:2013vb,Bernard:2018ivi,Bernard:2018hta} that encompass the scalar field effect upto the third PN order are converted in DJS conventions,\ i.e. the Einstein-Frame parameters (see, Table 1 of Ref.~\cite{Jain:2022nxs}).
The mass function $m(\varphi)$ is used to define the Einstein Frame parameters following Refs.~\cite{Damour:1992we,Damour:1995kt} \ i.e.
\begin{align}
\alpha_I&=\frac{d\ln m(\varphi)_I}{d\varphi},\\
\beta_I&=\frac{d\alpha_I}{d\varphi},\\
\beta'_I&=\frac{d\beta_I}{d\varphi},\\
\beta''_I&=\frac{d\beta'_I}{d\varphi}~.
\end{align} 

Finally, before proceeding to the computations of the scattering angle for ST theories, we briefly review the EOB formalism in ST theories. 
In the description of EOB, the relation between the real and EOB Hamiltonian is,
\begin{equation}
\label{eq:Hreal}
\hat{H}_{\rm real}\equiv \dfrac{H_{\rm real}}{\mu}=\dfrac{1}{\nu}\sqrt{1+2\nu\left(\hat{H}_{\rm eff}-1\right)},
\end{equation}
where $\nu=\mu/M$ is the symmetric mass ratio.  The reduced-mass effective Hamiltonian ($\hat{H}_{\text{eff}}$) is given by
\begin{align}
\label{heff-start}
\hat{H}_{\text{eff}}=\frac{H_{\text{eff}}}{\mu}=\sqrt{A(\hat{r})\left(1+\frac{\hat{p}_r^2}{B(\hat{r})}+\frac{\hat{p}_{\phi}^2}{\hat{r}^2}+\hat{Q}_e\right)}~,
\end{align}
where $\hat{p}_r$, $\hat{p}_{\phi}$ (with the magnitude $j=\hat{p}_{\phi}$) are the dimensionless radial and angular momenta, and $\hat{r}$ is the dimensionless radial separation. The dimensionless variables are defined as,
\begin{equation}
\hat{p}_{r}=\frac{P_{r}}{\mu};\quad\hat{p}_{\phi}=\frac{P_{\phi}}{G_{AB}M\mu};\quad\hat{r}=\frac{r}{G_{AB}M}~.
\end{equation}
Hereafter, the superscript \textit{hat} will be used to denote the dimensionless variables.

The three EOB potentials ($A, B, Q_e$) upto 3PN in gauge choice of Ref.~\cite{Damour:2000we} (also known as Damour, Jaranowski and Sch\"{a}fer gauge) formally read
\begin{align}
A(\hat{r})&=1-\frac{2}{\hat{r}}+\frac{a_2}{\hat{r}^2}+\frac{a_3}{\hat{r}^3}+\frac{a_4}{\hat{r}^4}~,\\
B(\hat{r})&=1+\frac{b_1}{\hat{r}}+\frac{b_2}{\hat{r}^2}+\frac{b_3}{\hat{r}^3}~,\\
\hat{Q}_e(\hat{r})&= q_3\frac{\hat{p}_r^4}{\hat{r}^2 }~.
\end{align}
where the $\nu$-dependent coefficients $a_i$, $b_i$ and $q_i$ take into account both GR and ST 
corrections which are separated as
\begin{align}
a_{i}&=a_{i}^{\rm GR}+ a_{i,\rm ST}~,\\
b_{i}&=b_{i}^{\rm GR}+b_{i,\rm ST},\\
q_{3}&=q_3^{\rm GR}+ q_{3, \rm ST}~.
\end{align}
The GR coefficients are known analytically up to 6PN \cite{Buonanno:1998gg,Damour:2000we,Bini:2020nsb,Bini:2020hmy,Damour:2015isa}, except for some unknown coefficients proportional to $\nu^2$. 
As for the ST theories, the nonlocal-in-time contributions start at the 3PN-order, the 3PN coefficients can be decomposed as 
\begin{align}
a_{4, \rm ST}&=a_{4, \rm ST}^{\rm I}+a_{4, \rm ST}^{\rm II}~,\\
b_{3, \rm ST}&=b_{3, \rm ST}^{\rm I}+b_{3, \rm ST}^{\rm II}~,\\
q_{3, \rm ST}&=q_{3, \rm ST}^{\rm I}+q_{3, \rm ST}^{\rm II}~,
\end{align}
where the superscripts $\rm{I}$ and $\rm{II}$ denote the local and non-local contributions, respectively. These coefficients can be further decomposed as,
\begin{align}
a_{4, \rm ST}^{\rm I}&=a_{\rm 4,ST}^{\rm loc}+a_{\rm 4, ST}^{\rm log}\ln(u),\\
b_{3, \rm ST}^{\rm I}&=b_{\rm 3,ST}^{\rm loc}+b_{\rm 3 ST}^{\rm log}\ln(u),\\
q_{3, \rm ST}^{\rm I}&=q_{\rm 3,ST}^{\rm loc}+q_{\rm 3, ST}^{\rm log}\ln(u)~.
\end{align}
These corrections to the EOB potentials in the ST theories upto 3PN order have been derived in \cite{Jain:2022nxs,Julie:2017pkb,Julie:2017ucp,Jain:2023fvt,Julie:2022qux}.
\section{Scalar-Tensor Scattering Angle : Local Contributions}
\label{Sec-locscat}
In this section, we derive the contribution to the scattering angle for encounters of two non-spinning bodies for the local part of the conservative dynamics up to third PN order in Scalar-Tensor (ST) theories. As the nonlocal-in-time (tail) effects start only at the 3PN-level in ST theories, the scattering angle upto 3PN can be separated as a sum of functions,
\begin{align}
\chi  = \chi_{\rm loc}+\chi_{\rm nonloc}~,
\end{align}
where $\chi_{\rm loc}$ and $\chi_{\rm nonloc}$ are local, nonlocal contributions to the scattering angle, respectively.

To compute the scattering angle we use the EOB-derived integral expression for the scattering angle $\chi$ \cite{Bini:2017wfr}. The action in EOB takes the form,
\begin{align}
S(T_{\rm eff},R,\phi;\mathcal{E}_{\rm eff},P_{\phi})=&-\mathcal{E}_{\rm eff}T_{\rm eff}+P_{\rm \phi}\phi\nonumber \\
&+\int^R dR~ P_{\rm R}(R,\mathcal{E}_{\rm eff},P_{\rm \phi})~,
\end{align}
where $\mathcal{E}_{\rm eff}$ is the EOB energy and $T_{\rm eff}$ is the EOB metric coordinate time. Using the equation of motion of orbit,
\begin{align}
\frac{\partial S}{\partial P_{\rm \phi}}=\phi_0=constant~,
\end{align}
and the Hamilton's equations, the scattering angle is
\begin{align}
\label{scat-form}
\frac{1}{2}\left(\chi (\bar{E},j)+\pi\right)=-\int_0^{u_{(max)}}\frac{1}{u^2}\frac{\partial}{\partial j}\hat{p}_r(\bar{E},j,u)~
\end{align}
where $u=1/\hat{r}$, $u_{(max)}=1/r_{min}$ is the distance of the closest approach of two bodies, and $\bar{E}$ is the dimensionless energy variable defined as \cite{Bini:2017wfr}
\begin{align}
\bar{E}\equiv\frac{1}{2}(\hat{\mathcal{E}}_{\rm eff}^2-1)\equiv\frac{1}{2}p_{\infty}^2~.
\end{align}

\subsection{PN-expanded $\chi_{\rm loc}$ for Scalar-Tensor theories}
Let us now first compute the radial momentum $\hat{p}_r$ as a function of $u=1/\hat{r}$, orbital angular momentum and energy which would then be used to compute the explicit integral of Eq.~\eqref{scat-form}. This is obtained by iteratively solving in $\hat{p}_r^2$ the EOB energy conservation law,
\begin{align}
\hat{\mathcal{E}}_{\rm eff}^2=A(u)\left(1+\frac{\hat{p}_r^2}{B(u)}+j^2 u^2+q_3~\hat{p}_r^4u^2\right)~,
\end{align}
which yields,
\begin{align}
\label{pr2eq}
\hat{p}_r^2=[\hat{p}_r^2]^0+[\hat{p}_r^2]^1\eta^2+[\hat{p}_r^2]^2\eta^4+[\hat{p}_r^2]^3\eta^6+\mathcal{O}(\eta^8)~
\end{align}
with $\eta\sim1/c$ as a PN-order marker\footnote{The explicit expressions of $\hat{p}_r^2$ and hence, $\hat{p}_r$ are given in the auxiliary file attached.}. 

This kind of formal PN expansion of $\hat{p}_r$ generates a sequence of divergent integral in the limit $[0,u_{(max)}]$. However, it was shown in \cite{Damour:1988mr} that the correct value of such a PN expanded integral is obtained by first using the Newtonian limit of $u_{(max)}$ as the upper limit of the integral and then taking the Hadamard partie finie of the integrals. The upper limit of the integral, $u_{max}$, is the positive root closest to zero of the Eq.~\eqref{pr2eq} and at the Newtonian level it reads
\begin{align}
u_{max}=\frac{1+\sqrt{1+2\bar{E}j^2}}{j^2}~.
\end{align} 

The integrals of Eq.~\eqref{scat-form} for ST theories are evaluated using this standard technique except a logarithmic integral arising at 3PN order. To simplify the expressions of the scattering angle, we introduce an auxiliary variable $\alpha$\footnote{Note that the parameter $\alpha$ defined here is different from the ST parameters $\alpha_I$, $\alpha(\varphi)$ and $\alpha_0$.}
\begin{equation}
\alpha=\frac{1}{\sqrt{\bar{E} j^2}}~,
\end{equation}
and a function
\begin{equation}
B(\alpha)=\rm{arctan(\alpha)}+\frac{\pi}{2}~.
\end{equation}
The scattering angle for local contribution $\frac{1}{2}\chi_{\rm loc}$ upto 3PN order can be decomposed as a sum of contributions from each PN order, \ i.e.
\begin{align}
\frac{1}{2}\chi_{\rm loc}^{\rm (N)}&=B(\alpha)-\frac{\pi}{2}\\ \nonumber\\
\frac{1}{2}\chi_{\rm loc}^{\rm (1PN)}&=\frac{1}{j^2}\left[C^{\rm B}_{(1 PN)} B(\alpha)+C^0_{\rm (1PN)}\right]\\ 
\frac{1}{2}\chi_{\rm loc}^{\rm (2PN)}&=\frac{1}{j^4}\left[C^{\rm B}_{\rm (2PN)}B(\alpha)+C^{0}_{\rm (2PN)}\right]\\
\label{3PN}
\frac{1}{2}\chi_{\rm loc}^{\rm (3PN)}&=\frac{1}{j^6}\left[C^{\rm B}_{(\rm 3PN)}B(\alpha)+C^{0}_{(\rm 3PN)}\right]+I_{\rm \chi}
\end{align}
where,
\begin{widetext}
\begin{align}
C^{\rm B}_{\rm (1PN)}&= \frac{1}{2} \left(-a_{\rm 2, ST}+b_{\rm 1, ST}+6\right)\nonumber \\
C^0_{(\rm 1PN)}&=\frac{1}{2\alpha \left(1+\alpha ^2 \right)}\left(6 \alpha ^2-\alpha ^2 a_{\rm 2, ST}+\alpha ^2 b_{\rm 1, ST}+b_{\rm 1, ST}+4\right)\nonumber\\
C^{\rm B}_{\rm (2PN)}&=\frac{1}{4\alpha^2}\left[105 \alpha ^2-3 \alpha ^2 a_{\rm 2, ST} b_{\rm 1, ST}-42 \alpha ^2 a_{\rm 2, ST}+\frac{3}{2} \alpha ^2 (a_{\rm 2, ST})^{\rm 2}-6 \alpha ^2 a_{\rm 3, ST}-2 a_{\rm 2, ST}+9 \alpha ^2 b_{\rm 1, ST}+3 \alpha ^2 b_{\rm 2, ST}\right.  \nonumber\\
& \left.-\frac{3}{4} \alpha ^2 b_{\rm 1, ST}^2+b_{\rm 1, ST}+b_{\rm 2, ST}-\frac{b_{\rm 1, ST}^2}{4}+15+\left(-30 \alpha ^2-6\right) \nu \right] \nonumber\\
C^{0}_{(\rm 2PN)}&=\frac{1}{4\alpha(1+\alpha^2)^2} \left[81 \alpha + 190 \alpha^3 + 105 \alpha^5-3 \alpha ^5 a_{\rm 2, ST} b_{\rm 1, ST}-5 \alpha ^3 a_{\rm 2, ST} b_{\rm 1, ST}-2 \alpha  a_{\rm 2, ST} b_{\rm 1, ST}\right.\nonumber \\
&\left.-42 \alpha ^5 a_{\rm 2, ST}+\frac{3}{2} \alpha ^5 a_{\rm 2, ST}^2-6 \alpha ^5 a_{\rm 3, ST}-72 \alpha ^3 a_{\rm 2, ST}+\frac{5}{2} \alpha ^3 a_{\rm 2, ST}^2-10 \alpha ^3 a_{\rm 3, ST}-26 \alpha  a_{\rm 2, ST}\right.\nonumber \\
&\left.-4 \alpha  a_{\rm 3, ST}+9 \alpha ^5 b_{\rm 1, ST}+3 \alpha ^5 b_{\rm 2, ST}-\frac{3}{4} \alpha ^5 b_{\rm 1, ST}^2+16 \alpha ^3 b_{\rm 1, ST}+6 \alpha ^3 b_{\rm 2, ST}-\frac{3}{2} \alpha ^3 b_{\rm 1, ST}^2+7 \alpha  b_{\rm 1, ST}\right.\nonumber \\
&\left.+3 \alpha  b_{\rm 2, ST}-\frac{3}{4} \alpha  b_{\rm 1, ST}^2+\left(-30 \alpha ^5-56 \alpha ^3-26 \alpha \right) \nu\right]\nonumber\\
C^{\rm B}_{(\rm 3PN)}&=\frac{1}{\alpha^4}\left[\frac{9 \nu ^2}{8}-\frac{3 \nu }{2}-\frac{3 q_{\rm 3, ST}}{16}\right]+\frac{1}{\alpha^2}\left[\frac{3}{32} a_{\rm 2, ST} b_{\rm 1, ST}^2-\frac{21 a_{\rm 2, ST} b_{\rm 1, ST}}{8}-\frac{3 a_{\rm 2, ST} b_{\rm 2, ST}}{8}+\frac{9 \nu  a_{\rm 2, ST}}{4}+\frac{3 a_{\rm 2, ST}^2}{2}\right.\nonumber \\
&\left.-\frac{273 a_{\rm 2, ST}}{8}-\frac{3 a_{\rm 3, ST}b_{\rm 1, ST}}{8}-\frac{27 a_{\rm 3, ST}}{4}-\frac{3 a_{\rm 4, ST}^{\rm loc}}{4}-\frac{3 b_{\rm 1, ST} b_{\rm 2, ST}}{8}+\frac{3 \nu  b_{\rm 1, ST}}{2}+\frac{3 b_{\rm 1, ST}^3}{32}-\frac{3 b_{\rm 1, ST}^2}{16}+\frac{45 b_{\rm 1, ST}}{8}\right.\nonumber \\
&\left.+\frac{9 b_{\rm 2, ST}}{4}+\frac{3b_{\rm 3, ST}^{\rm loc}}{4}+\frac{45 \nu ^2}{4}+\frac{123 \pi ^2 \nu }{128}-109 \nu -\frac{9 q_{\rm 3, ST}^{\rm loc}}{8}+\frac{315}{4}\right]+\frac{15 a_{\rm 2, ST} a_{\rm 3, ST}}{4}+\frac{15}{16} a_{\rm 2, ST}^2 b_{\rm 1, ST}\nonumber \\
&+\frac{15}{32} a_{\rm 2, ST} b_{\rm 1, ST}^2-\frac{105 a_{\rm 2, ST} b_{\rm 1, ST}}{8}-\frac{15 a_{\rm 2, ST} b_{\rm 2, ST}}{8}+\frac{75 \nu  a_{\rm 2, ST}}{4}-\frac{1}{16} 5 a_{\rm 2, ST}^3+\frac{195 a_{\rm 2, ST}^2}{8}-\frac{1485 a_{\rm 2, ST}}{8}\nonumber \\
&-\frac{15 a_{\rm 3, ST}b_{\rm 1, ST}}{8}-\frac{135 a_{\rm 3, ST}}{4}-\frac{15 a_{\rm 4, ST}^{\rm loc}}{4}-\frac{5 b_{\rm 1, ST} b_{\rm 2, ST}}{8}+\frac{5 b_{\rm 1, ST}^3}{32}+\frac{175 b_{\rm 1, ST}}{8}-\frac{15 b_{\rm 1, ST}^2}{16}+\frac{25 b_{\rm 2, ST}}{4}\nonumber \\
&+\frac{5b_{\rm 3, ST}^{\rm loc}}{4}+\frac{105 \nu ^2}{8}+\frac{615 \pi ^2 \nu }{128}-\frac{625 \nu }{2}-\frac{15 q_{\rm 3, ST}^{\rm loc}}{16}+\frac{1155}{4}\nonumber\\
C^{0}_{(\rm 3PN)}&=\frac{1}{(1+\alpha^2)^3}\left\{\alpha^5\left[\frac{15 a_{\rm 2, ST}a_{\rm 3, ST}}{4}+\frac{15}{16} a_{\rm 2, ST}^2 b_{\rm 1, ST}+\frac{15}{32} a_{\rm 2, ST}b_{\rm 1, ST}^2-\frac{105 a_{\rm 2, ST}b_{\rm 1, ST}}{8}-\frac{15 a_{\rm 2, ST}b_{\rm 2, ST}}{8}\right.\right.\nonumber\\
&\left.\left.+\nu  \left(\frac{75 a_{\rm 2, ST}}{4}+\frac{615 \pi ^2}{128}-\frac{625}{2}\right)-\frac{1}{16} 5 a_{\rm 2, ST}^3+\frac{195 a_{\rm 2, ST}^2}{8}-\frac{1485 a_{\rm 2, ST}}{8}-\frac{15 a_{\rm 3, ST}b_{\rm 1, ST}}{8}-\frac{135 a_{\rm 3, ST}}{4}\right.\right.\nonumber\\
&\left.\left.-\frac{15 a_{\rm 4, ST}^{\rm loc}}{4}-\frac{5 b_{\rm 1, ST}b_{\rm 2, ST}}{8}+\frac{5 b_{\rm 1, ST}^3}{32}+\frac{175 b_{\rm 1, ST}}{8}-\frac{15 b_{\rm 1, ST}^2}{16}+\frac{25 b_{\rm 2, ST}}{4}+\frac{5b_{\rm 3, ST}^{\rm loc}}{4}+\frac{105 \nu ^2}{8}-\frac{15 q_{\rm 3, ST}^{\rm loc}}{16}\right.\right.\nonumber\\
&\left.\left.+\frac{1155}{4}\right]+\alpha^3\left[10 a_{\rm 2, ST} a_{\rm 3, ST}+\nu  \left(\frac{209 a_{\rm 2, ST}}{4}+\frac{3 b_{\rm 1, ST}}{2}+\frac{1763 \pi ^2}{128}-\frac{2827}{3}\right)+\frac{5}{2} a_{\rm 2, ST}^2 b_{\rm 1, ST}+\frac{43}{32} a_{\rm 2, ST} b_{\rm 1, ST}^2\right.\right.\nonumber\\
&\left.\left.-\frac{301 a_{\rm 2, ST} b_{\rm 1, ST}}{8}-\frac{43 a_{\rm 2, ST} b_{\rm 2, ST}}{8}-\frac{1}{6} 5 a_{\rm 2, ST}^3+\frac{133 a_{\rm 2, ST}^2}{2}-\frac{4233 a_{\rm 2, ST}}{8}-\frac{43 a_{\rm 3, ST} b_{\rm 1, ST}}{8}-\frac{387 a_{\rm 3, ST}}{4}-\frac{43 a_{\rm 4, ST}^{\rm loc}}{4}\right.\right.\nonumber\\
&\left.\left.-\frac{49 b_{\rm 1, ST} b_{\rm 2, ST}}{24}+\frac{49 b_{\rm 1, ST}^3}{96}+\frac{1535 b_{\rm 1, ST}}{24}-\frac{43 b_{\rm 1, ST}^2}{16}+\frac{227 b_{\rm 2, ST}}{12}+\frac{49 b_{\rm 3, ST}^{\rm loc}}{12}+\frac{185 \nu ^2}{4}-\frac{29 q_{\rm 3, ST}^{\rm loc}}{8}+\frac{3395}{4}\right]\right.\nonumber\\
&\left.+\alpha\left[\frac{33 a_{\rm 2, ST} a_{\rm 3, ST}}{4}+\nu  \left(\frac{189 a_{\rm 2, ST}}{4}+4 b_{\rm 1, ST}+\frac{1681 \pi ^2}{128}-\frac{2939}{3}\right)+\frac{33}{16} a_{\rm 2, ST}^2 b_{\rm 1, ST}+\frac{41}{32} a_{\rm 2, ST} b_{\rm 1, ST}^2\right.\right.\nonumber\\
&\left.\left.-\frac{287 a_{\rm 2, ST} b_{\rm 1, ST}}{8}-\frac{41 a_{\rm 2, ST} b_{\rm 2, ST}}{8}-\frac{1}{16} 11 a_{\rm 2, ST}^3+\frac{461 a_{\rm 2, ST}^2}{8}-\frac{3995 a_{\rm 2, ST}}{8}-\frac{41 a_{\rm 3, ST} b_{\rm 1, ST}}{8}-\frac{369 a_{\rm 3, ST}}{4}-\frac{41 a_{\rm 4, ST}^{\rm loc}}{4}\right.\right.\nonumber\\
&\left.\left.-\frac{19 b_{\rm 1, ST} b_{\rm 2, ST}}{8}+\frac{19 b_{\rm 1, ST}^3}{32}+\frac{505 b_{\rm 1, ST}}{8}-\frac{41 b_{\rm 1, ST}^2}{16}+\frac{79 b_{\rm 2, ST}}{4}+\frac{19 b_{\rm 3, ST}^{\rm loc}}{4}+60 \nu ^2-\frac{21 q_{\rm 3, ST}^{\rm loc}}{4}+\frac{3381}{4}\right]\right.\nonumber\\
&\left.+\frac{1}{\alpha}\left[2 a_{\rm 2, ST} a_{\rm 3, ST}+\nu  \left(\frac{55 a_{\rm 2, ST}}{4}+\frac{7 b_{\rm 1, ST}}{2}+\frac{533 \pi ^2}{128}-\frac{1153}{3}\right)+\frac{13}{32} a_{\rm 2, ST} b_{\rm 1, ST}^2+\frac{1}{2} a_{\rm 2, ST}^2 b_{\rm 1, ST}-\frac{95 a_{\rm 2, ST} b_{\rm 1, ST}}{8}\right.\right.\nonumber\\
&\left.\left.-\frac{13 a_{\rm 2, ST} b_{\rm 2, ST}}{8}+\frac{29 a_{\rm 2, ST}^2}{2}-\frac{1279 a_{\rm 2, ST}}{8}-\frac{13 a_{\rm 3, ST} b_{\rm 1, ST}}{8}-\frac{121 a_{\rm 3, ST}}{4}-\frac{13 a_{\rm 4, ST}^{\rm loc}}{4}-\frac{9 b_{\rm 1, ST} b_{\rm 2, ST}}{8}\right.\right.\nonumber\\
&\left.\left.+\frac{9 b_{\rm 1, ST}^3}{32}-\frac{13 b_{\rm 1, ST}^2}{16}+\frac{179 b_{\rm 1, ST}}{8}+\frac{31 b_{\rm 2, ST}}{4}+\frac{9 b_{\rm 3, ST}^{\rm loc}}{4}+\frac{135 \nu ^2}{4}-\frac{27 q_{\rm 3, ST}^{\rm loc}}{8}+\frac{1221}{4}\right]\right.\nonumber\\
&\left.+\frac{1}{\alpha^3}\left[-\frac{a_{\rm 2, ST} b_{\rm 1, ST}}{2}-6 a_{\rm 2, ST}-a_{\rm 3, ST}-\frac{b_{\rm 1, ST} b_{\rm 2, ST}}{6}+\nu  \left(b_{\rm 1, ST}-\frac{69}{2}\right)+\frac{b_{\rm 1, ST}^3}{24}+\frac{4 b_{\rm 1, ST}}{3}+\frac{2 b_{\rm 2, ST}}{3}+\frac{b_{\rm 3, ST}^{\rm loc}}{3}\right.\right.\nonumber\\
&\left.\left.+\frac{55 \nu ^2}{8}-\frac{13 q_{\rm 3, ST}^{\rm loc}}{16}+\frac{64}{3}\right]\right\}
\end{align}
\end{widetext}

Here,  for simplicity we do not substitute the values of the ST corrections $a_{i,\rm ST}$, $b_{i,\rm ST}$ and $q_{i,\rm ST}$. The explicit expressions of the corrections have been derived in Refs.~\cite{Julie:2017pkb,Julie:2017ucp,Jain:2022nxs,Jain:2023fvt,Julie:2022qux}.

Finally, the last contribution, $I_{\rm \chi}$, to the 3PN scattering angle is,
\begin{align}
\label{3PN-ini}
I_{\rm \chi}&=-\frac{j a_{\rm 4,ST}^{\rm log}}{2}\mathrm{Pf}\int_{0}^{u_{\rm max}}\frac{ u^4 \ln (u)}{\left(2 \bar{E}+2 u -j^2 u^2\right)^{3/2}} du~.
\end{align}
Since this integral can not be solved using the standard techniques as above, we simplify the integral by using suitable integration by parts as
\begin{align}
\label{3PN-log}
I_{\rm \chi}=&a_{\rm 4,log}^{\rm ST}\frac{(15\alpha^4 + 18\alpha^2 + 3)}{16j^6\alpha^2(\alpha^2 + 1)} B(\alpha)\nonumber \\
&+ a_{\rm 4,ST}^{\rm log}\frac{(15\alpha^2 + 13)}{16(\alpha^2 + 1)\alpha j^6}+\mathcal{I}_{\rm \chi}~,
\end{align}
where the last term is now a \textit{convergent} integral defined as
\begin{align}
\label{int2}
\mathcal{I}_{\chi}=\frac{2 j a_{\rm 4,ST}^{\rm log}}{(1+2 j^2\bar{E})}\int_{0}^{u_{max}}\frac{u^3 (u j^2 - 1)\ln(u)}{\sqrt{ 2\bar{E} + 2u-j^2u^2}}du~.
\end{align}
The integral of Eq.~\eqref{int2} can not be expressed in terms of the elementary functions.  However, after suitable change of variables,
\begin{align}
u=\frac{\sqrt{2\bar{E}}}{j} x\quad;\quad\epsilon\equiv2\alpha=\frac{2}{p_{\infty}j}~,
\end{align}
the integral can be computed in large $j$-expansion (small $\epsilon$-expansion) at fixed $p_{\infty}$. We follow the approach of Ref.~\cite{Bini:2017wfr} to compute the $j$-expansion of the integral. Here, we display the first three contributions to integral in $j$-expansion,
\begin{align}
\label{3PN-logfi}
\mathcal{I}_{\chi}&=\frac{a_{\rm 4, ST}^{\rm log}\bar{E}}{j^4}\mathcal{I}_4+\frac{a_{\rm 4, ST}^{\rm log}\bar{E}^{1/2}}{j^5}\mathcal{I}_5+\frac{a_{\rm 4, ST}^{\rm log}}{j^6}\mathcal{I}_6+\mathcal{O}\left(\frac{\bar{E}^{-1/2}}{j^7}\right)~,
\end{align}
where 
\begin{align}
\mathcal{I}_4&=\frac{\pi}{16}\left[7+6 \ln\left(\frac{\bar{E}}{2j^2}\right)\right]~,\\ \nonumber
\mathcal{I}_5&=\sqrt{2}\left[-2+8\ln(2)+2\ln\left(\frac{\bar{E}}{2j^2}\right)\right]~,\\ \nonumber
\mathcal{I}_6&=\frac{\pi}{32}\left[77+30\ln\left(\frac{\bar{E}}{2j^2}\right)\right]~.\\ \nonumber
\end{align}
The higher order contributions can be computed following the same approach.
\newline
\subsection{Final result of the 3PN scattering angle in large $j$-expansion}
The result presented in Eq.~\eqref{3PN} of the scattering angle at the 3PN-order is fully explicit except the integral $I_{\rm \chi}$ of Eq.~\eqref{3PN-ini}.  To compute this integral, we expressed it into a simpler integral of Eq.~\eqref{3PN-log}. Then at the end of the last subsection, we computed large $j$-expansion of this remaining part, $\mathcal{I}_{\rm \chi}$,  of the integral $I_{\rm \chi}$.

Let us now insert the results of Eq.~\eqref{3PN-log} in the large $j$-expansion of the scattering angle at the 3PN order. As the contributions to both $I_{\rm \chi}$ and $\mathcal{I}_{\rm \chi}$ start at $1/j^4$ order in their large $j$-expansion, we only show the large $j$-expansion of the exactly known part of $\chi_{\rm loc}^{\rm (3PN)}/2$ upto $1/j^4$, \ i.e.
\begin{widetext}
\begin{align}
\label{largej1}
\frac{\chi_{\rm 3PN}}{2}-\mathcal{I}_{\chi}&=\frac{p_{\infty}^4}{j^2}\left[-\frac{3}{32} \pi   q_{\rm 3, ST}^{\rm loc}+\frac{9}{16} \pi  \nu ^2 -\frac{3}{4} \pi  \nu  \right]+\frac{p_{\infty}^3}{j^3}\left[-\frac{a_{\rm 2, ST}b_{\rm 1, ST}}{2}-6 a_{\rm 2,ST}-a_{\rm 3,ST}-\frac{b_{\rm 1, ST}b_{\rm 2, ST}}{6}+\nu  b_{\rm 1, ST}+\frac{b_{\rm 1, ST}^3}{24}\right.\nonumber\\
&\left.+\frac{4 b_{\rm 1, ST}}{3}+\frac{2 b_{\rm 2, ST}}{3}+\frac{b_{\rm 3, ST}^{\rm loc}}{3}+8 \nu ^2-36 \nu -q_{\rm 3, ST}^{\rm loc}+\frac{64}{3}\right]+\frac{p_{\infty}^2}{j^4}\left[\frac{3}{64} \pi  a_{\rm 2, ST}b_{\rm 1, ST}^2-\frac{21}{16} \pi  a_{\rm 2, ST}b_{\rm 1, ST}\right.\nonumber\\
&\left.-\frac{3}{16} \pi  a_{\rm 2, ST}b_{\rm 2, ST}+\frac{9}{8} \pi  \nu  a_{\rm 2,ST}+\frac{3 \pi  a_{\rm 2,ST}^2}{4}-\frac{273 \pi  a_{\rm 2,ST}}{16}-\frac{3}{16} \pi  a_{\rm 3, ST}b_{\rm 1, ST}-\frac{27 \pi  a_{\rm 3,ST}}{8}-\frac{3 \pi  a_{\rm 4,ST}^{\rm loc}}{8}+\frac{3 \pi  a_{\rm 4, ST}^{\rm log}}{32}\right.\nonumber \\
&\left.-\frac{3}{16} \pi  b_{\rm 1, ST}b_{\rm 2, ST}+\frac{3}{4} \pi  \nu  b_{\rm 1, ST}+\frac{3 \pi  b_{\rm 1, ST}^3}{64}-\frac{3 \pi  b_{\rm 1, ST}^2}{32}+\frac{45 \pi  b_{\rm 1, ST}}{16}+\frac{9 \pi  b_{\rm 2, ST}}{8}+\frac{3 \pi  b_{\rm 3, ST}^{\rm loc}}{8}+\frac{45 \pi  \nu ^2}{8}+\frac{123 \pi ^3 \nu }{256}\right.\nonumber \\
&\left.-\frac{109 \pi  \nu }{2}-\frac{9 \pi  q_{\rm 3, ST}^{\rm loc}}{16}+\frac{315 \pi }{8}\right]+\mathcal{O}\left(\frac{1}{j^5}\right)
\end{align}
\end{widetext}
In Eq.~\eqref{3PN-logfi}, we computed the integral $\mathcal{I}_{\rm \chi}$ in the large $j$-expansion, and its first term reads,
\begin{align}
\label{largej2}
\mathcal{I}_{\chi}=\frac{\pi a_{\rm 4, ST}^{\rm log} p_{\infty}^2}{32j^4}\left[7+6 \ln\left(\frac{\bar{E}}{2j^2}\right)\right] +\mathcal{O}\left(\frac{1}{j^5}\right)~.
\end{align}
Now, combining Eqs.~(\ref{largej1}) and (\ref{largej2}) gives the large $j$-expansion of the total scattering angle at the 3PN order for local part of the dynamics in ST theories. 

\section{Non-local contributions to the scattering angle}
\label{Sec-nonlocscat}
In this section, we compute the leading order (LO) nonlocal contributions to the scattering angle using the order-reduction approach of Ref.~\cite{Damour:2015isa} for bound orbits. This approach has been recently used to derive the nonlocal contributions to the EOB metric potentials for bound orbits in ST theories \cite{Jain:2023fvt,Julie:2022qux}. Here, we will use this approach for hyperboliclike orbits in ST theories following Refs.~\cite{Bini:2017wfr,Bini:2021jmj}.

As the tail contribution to the Hamiltonian starts at 3PN order in ST theory one can compute the scattering angle $\chi_{\rm nonloc}$ by considering the Hamiltonian
\begin{align}
H=H_{\rm N} +H^{\rm tail}~,
\end{align}
where $H_{\rm N}$ is the Newtonian-order Hamiltonian and $H^{\rm tail}$ is the LO tail contribution \cite{Bernard:2018ivi}, as all the other PN contributions upto 3PN order have been already considered in local scattering angle computation.

For the computation of the scattering angle using the general Hamilton-Jacobi derived equation,
\begin{equation}
\label{chihj}
\chi(\bar{E},j)=-\frac{\partial}{\partial j}\int \hat{p}_{r}(\bar{E},j,r) dr~,
\end{equation}
the function radial momentum is first computed by solving for $\hat{p}_r^2$ the energy conservation law,
\begin{align}
\bar{E}=\frac{H_{\rm real}(\hat{r},\hat{p}_r,j)}{\nu}=\frac{1}{2}(\hat{p}_r^2+\frac{j^2}{\hat{r}^2})-\frac{1}{\hat{r}}+\frac{H^{\rm tail}}{\nu}~,
\end{align}
in $M=1$ ($\mu=\nu$) units.  At LO in tail, the solution of the equation in $\hat{p}_r$ is
\begin{align}
\hat{p}_r=\hat{p}_r^{0}-\frac{1}{\hat{p}_r^{0}}\frac{H^{\rm tail}(\hat{r},\hat{p}_r^{0},j)}{\nu}~,
\end{align}
where $\hat{p}_r^0$  is the Newtonian contribution. Inserting the solution in Eq.~\eqref{chihj}, the nonlocal contribution to the scattering angle reads
\begin{align}
\chi_{\rm nonloc}=\frac{1}{\nu}\frac{\partial}{\partial j}W^{\rm tail}(\bar{E},j)~,
\end{align}
where 
\begin{align}
W^{\rm tail}&=\int\frac{dr}{\hat{p}_r^0}H^{\rm tail}(\hat{r},\hat{p}_r,j)\nonumber\\
&=\int dt H^{\rm tail}~.
\end{align}

The LO tail contribution to the Hamiltonian in ST theories reads \cite{Bernard:2018ivi}
\begin{align}
H^{\rm tail}_{\rm LO}=-\frac{2 G^2 }{3c^6}(3+2w_0)\mathrm{Pf}_{\rm 2s/c}\int_{-\infty}^{\infty}\frac{d\tau}{|\tau|} I_{\rm s,i}^{(2)}(t) I_{s,i}^{(2)}(t+\tau)~,
\end{align}
where $\mathrm{Pf}$ is the Hadamard partie finie function with the Hadamard partie finie scale $s$ and $I_{\rm s,i}^{(2)}$ is the second time-derivative of the scalar dipole moment, $I_{\rm s,i}$. In the center-of-mass (COM) frame, the scalar dipole moment is
\begin{align}
\label{scalardm}
I_{\rm s,i}=\frac{2 \mu \alpha_0^2}{\phi_0} (s_A-s_B) x^i~,
\end{align}
where $s_{A}$, $s_B$ are the sensitivities of two bodies. Thus the LO potential $W^{\rm tail}$ is,
\begin{align}
\label{Wtail}
W^{\rm tail}&=\int dt~H^{\rm tail}_{\rm LO}\nonumber \\
&=-\frac{2 G^2 }{3c^6}(3+2w_0)\int dt\mathrm{Pf}_{\rm 2s/c}\int_{-\infty}^{\infty}\frac{d\tau}{|\tau|} I_{\rm s,i}^{(2)}(t) I_{s,i}^{(2)}(t+\tau)~.
\end{align}

In Refs.~\cite{Jain:2023fvt,Julie:2022qux}, it is shown that the scalar dipole moment in action-angle variables and using the Kepler's equations for ellipticlike orbits is a periodic function, and hence can be decomposed into Fourier series. Here, we are considering hyperboliclike motions, therefore the Cartesian coordinates are parameterised as
\begin{align}
\label{carthp}
x =& -a(\mathrm{cosh}\bar{u}-e)~,\\
y=&-a\sqrt{e^2-1}~\mathrm{sinh}\bar{u}~,
\end{align}
and the hyperbolic Kepler equation is
\begin{align}
\label{carthp1}
\bar{n}t=e~\mathrm{sinh}\bar{u}-\bar{u}~,
\end{align}
where $e$ is the eccentricity, $a$ is the semi-major axis, and \begin{align}
\bar{n}=\frac{1}{\bar{a}^{3/2}}\quad;\quad \bar{a}=-a~.
\end{align}
Similar to the ellipticlike orbits, the scalar dipole moment for hyperboliclike motions can also be decomposed into Fourier series,\ i.e.
\begin{align}
I_{\rm s,i}&=\int \frac{d\omega}{2\pi} \tilde{I}_{s,i}(\omega) \mathrm{e}^{-i\omega t}~,\\ \nonumber
\tilde{I}_{\rm s,i}&=\int dt~ I_{\rm s,i} \mathrm{e}^{i\omega t}~,
\end{align}
where $\tilde{I}_{s,i}(\omega)$ is the Fourier-transform of the scalar dipole moment.

Inserting the Fourier transformation of the scalar dipole moment (and $\tau=G_{AB}M(t'-t)$)\footnote{Here, the dimensionless variable $\hat{\tau}=\frac{\tau}{G_{AB}}$ and $t=\frac{T}{G_{AB}M}$.} in Eq.~(\ref{Wtail}) yields,
\begin{widetext}
\begin{align}
\label{pot}
W_{\rm tail}&=-\frac{1}{(G_{AB})^2}\frac{2G^2}{3}(3+2w_0)\int dt~ \mathrm{Pf}_{2\hat{s}/c}\int\frac{dt'}{|t-t'|}\int\frac{d\omega}{2\pi}\frac{d\omega'}{2\pi}\omega^2\omega'^2\tilde{I}_{s,i}(\omega)\tilde{I}_{s,i}(\omega')\mathrm{e}^{-i\omega t}\mathrm{e}^{-i\omega' t'}\nonumber \\
&=-\frac{2G^2}{3G_{AB}^2}(3+2w_0)\mathrm{Pf}_{\rm 2\hat{s}/c}\int\frac{dt'}{|t-t'|}\int \frac{d\omega}{2\pi}\frac{d\omega'}{2\pi}\omega^2\omega'^2\tilde{I}_{\rm s,i}(\omega)\tilde{I}_{\rm s,i}(\omega')\mathrm{e}^{-i\omega' \hat{\tau}}~2\pi\delta(\omega+\omega')
\nonumber\\
&=-\frac{2G^2}{3G_{AB}^2}(3+2w_0)\mathrm{Pf}_{\rm 2\hat{s}/c}\int\frac{dt'}{|t-t'|}\int \frac{d\omega}{2\pi}\omega^4\tilde{I}_{\rm s,i}(\omega)\tilde{I}_{\rm s,i}(-\omega)\mathrm{e}^{i\omega \hat{\tau}}\nonumber\\
&=-\frac{2G^2}{3G_{AB}^2}(3+2w_0)\int \frac{d\omega}{2\pi}\omega^4\tilde{I}_{\rm s,i}(\omega)\tilde{I}_{\rm s,i}(-\omega)\mathrm{Pf}_{\rm 2\hat{s}/c}\int\frac{d\hat{\tau}}{|\hat{\tau}|}\mathrm{e}^{i\omega \hat{\tau}}
\end{align}
\end{widetext}
The partie finie integral of the last term in the above equation is,
\begin{align}
\label{partiefinie}
\mathrm{Pf}_{2\hat{s}/c}\int_{-\infty}^{\infty}\frac{d\tau}{|\tau|}\mathrm{e}^{i\omega \tau}=-2\ln\left(\frac{2|\omega|\hat{s}~\mathrm{e}^{\gamma_{\rm Euler}}}{c}\right)~,
\end{align}
where $\gamma_{\rm Euler}$ is Euler's Gamma. Inserting Eq.~\eqref{partiefinie} into Eq.~\eqref{pot} gives the Fourier-domain formula for the potential,
\begin{align}
\label{potf}
W^{\rm tail}=\frac{8G^2}{3G_{AB}^2}(3+2w_0)\int_{0}^{\infty} &\frac{d\omega}{2\pi}\omega^4\tilde{I}_{\rm s,i}(\omega)\tilde{I}_{\rm s,i}^{*}(\omega)\nonumber\\&\ln\left(\frac{2|\omega|\hat{s}~\mathrm{e}^{\gamma_{\rm Euler}}}{c}\right)~,
\end{align}
where $\tilde{I}_{\rm s,i}(-\omega)=\tilde{I}_{\rm s,i}^{\ast}(\omega)$.

To compute the explicit expression of Eq.~\eqref{potf} of potential $W^{rm tail}$ in terms of $\bar{E}$ and $j$, we insert the Fourier transform of scalar dipole moment.  For this, we evaluate the Fourier transforms of $(x,y)$ for hyperbolic orbits,\ i.e
\begin{align}
\label{fthyp}
x&=\int dt~ \mathrm{e}^{i\omega t}x(t)~,\nonumber \\
y&=\int dt ~\mathrm{e}^{i\omega t}y(t)~. 
\end{align}

Inserting Eqs.~\eqref{carthp}-\eqref{carthp1} into Eq.~\eqref{fthyp} and using the definition of Hankel functions of first kind (see Eq.~(9.1.25) of \cite{10.5555/1098650}), 
\begin{align}
\int^{\infty}_{-\infty}\mathrm{e}^{q \mathrm{sinh}\xi -p\xi}=i\pi H^{(1)}_p(q)~,
\end{align}
we find the Fourier transform as
\begin{align}
\label{ftx}
x&=\frac{\pi a}{\omega}\left(\frac{p}{q}H^{(1)}_p(q)-H^{(1)}_{p+1}(q)\right)~,\\
\label{fty}
y&=-\frac{\pi a}{\omega e}\sqrt{e^2-1}~H^{(1)}_p(q)~
\end{align}
where
\begin{align}
q=i e \frac{\omega}{\bar{n}}~;\quad p=\frac{q}{e}~.
\end{align}

We then consider the Fourier transform of $(x,y)$ in large-$j$ limit which is equivalent to large-$e$ limit $e=\sqrt{1+2\bar{E}j^2}$. The large-$e$ limit of Eqs.~\eqref{ftx}-\eqref{fty}  yields,
\begin{align}
\label{ftx1}
x&=-\frac{\pi a}{\omega}~H^{(1)}_{1}(iu)~,\\
\label{fty1}
y&=-\frac{\pi a}{\omega }~H^{(1)}_0(iu)~
\end{align}
where $q=iu$. The Hankel functions evaluated at purely imaginary arguments are related to modified Bessel functions $K_\nu$ as (see Eq.~(9.6.4) of \cite{10.5555/1098650})
\begin{align}
\label{Htob}
K_0(x)=i\frac{\pi}{2}H^{(1)}_0(ix)~;\quad K_1(x)=-\frac{\pi}{2}H^{(1)}_1(ix)~.
\end{align}

Finally, inserting Eqs.~\eqref{ftx1}-\eqref{Htob} in Eq.~\eqref{potf} and then taking $j$-derivative of potential $W^{\rm tail}$, the explicit expression of the scattering angle in large-$e$ limit yields,
\begin{align}
\label{scattail}
\chi^{\rm tail}_{\rm 3PN}=-\frac{2\pi \nu}{3}\frac{p_{\infty}^2}{j^4}&\left[2\delta_+ +\frac{\bar{\gamma}_{AB}(\bar{\gamma}_{AB}+2)}{2}\right]\nonumber\\&\left\{7+3\ln\left(\frac{p_{\infty}^2\hat{s}}{4 j }\right)\right\}~.
\end{align}
where we recall that $\hat{s}=s/(G_{AB}M)$ is the dimensionless regularisation scale defining the nearzone-farzone separation. Here, the subscript "$\pm$" denotes the symmetric and anti-symmetric parts of the ST parameters, e.g. $z_{\pm}=(z_A \pm z_B)/2$.

\section{Summing the local and non-local contributions to $\chi_{\rm 3PN}$ in large-j expansion}
\label{Sec-sum3PN}
In Sec.~\ref{Sec-locscat}, we first computed the local scattering angle upto 3PN order and then in Sec.~\ref{Sec-nonlocscat} we separately computed the nonlocal contributions at the 3PN order. The results at 1PN and 2PN levels were given in fully explicit and exact form. However, the results at the 3PN order were obtained in the large-$j$ expansion for both the local contribution (due to the logarithmic term $\mathcal{I}_{\rm \chi}$) and the nonlocal contribution. On combining the two separate 3PN order contributions to the scattering angle at 3PN, we find
\begin{widetext}
\begin{align}
\frac{\chi(\bar{E},j)^{\rm (3PN)}}{2}&=\frac{p_{\infty}^4}{j^2}\pi\left\{\nu  \left[-\frac{3}{4}-\frac{15}{64}   \bar{\gamma }_{\text{AB}}^2-\frac{13}{16} \bar{\gamma }_{\text{AB}}-\frac{1 }{16}\langle \bar{\beta}\rangle+\frac{1}{16}\langle\delta\rangle\right]+\nu ^2 \left[\frac{3}{8} \bar{\gamma }_{\text{AB}}-\frac{3}{16}    \langle\bar{\beta}\rangle+\frac{9  }{16}\right]\right\}\nonumber\\&+\frac{p_{\infty}^3}{j^3}\left\{\frac{64}{3}-\frac{4}{3} \left[\langle \delta \rangle \left(\bar{\gamma }_{\text{AB}}+2\right)-2 \bar{\gamma }_{\text{AB}} \left(\bar{\gamma }_{\text{AB}}^2+6 \bar{\gamma }_{\text{AB}}+12\right)\right]+\frac{\nu}{6}\left[-216+4 \langle\bar{\beta }\rangle \left(8 \bar{\gamma }_{\text{AB}}-3\right)+8\bar{\gamma }_{\text{AB}} \delta _+ -6 \bar{\gamma }_{\text{AB}}^3\right.\right.\nonumber\\
&\left.\left.-93 \bar{\gamma }_{\text{AB}}^2-262 \bar{\gamma }_{\text{AB}}-4 X_{\text{AB}} \bar{\gamma }_{\text{AB}}\beta _- +8 \langle \delta \rangle-57  X_{\text{AB}}\beta _-+2  X_{\text{AB}}\epsilon _-+36 \beta _++4 \delta _+-2 \zeta -2 \epsilon _+\right]+\frac{\nu^2}{3} \left[\bar{\gamma }_{\text{AB}}^2+22 \bar{\gamma }_{\text{AB}}\right.\right.\nonumber\\
&\left.\left.-6 \langle\bar{\beta }\rangle-18 \beta _++4 \delta _+-6 \zeta +2 \epsilon _++24\right]\right\}\nonumber\\
&+\frac{p_{\infty}^2}{j^4}\pi\left\{\frac{315}{8}+\frac{1}{32}  \left[4 \langle \delta \rangle \left(-20 \bar{\gamma }_{\text{AB}}+12 \beta _+^3+12 \beta _-^2 \beta _+-35\right)+\langle\bar{\beta }\rangle \left(-187 \bar{\gamma }_{\text{AB}}^2-556 \bar{\gamma }_{\text{AB}}+12 \langle \delta \rangle-412\right)+16  \bar{\gamma }_{\text{AB}}\langle \epsilon \rangle\right.\right.\nonumber\\
&\left.\left.+236 \bar{\gamma }_{\text{AB}}^3+1229 \bar{\gamma }_{\text{AB}}^2+2148 \bar{\gamma }_{\text{AB}}+48 \langle\bar{\beta }\rangle^2-8 \langle \kappa \rangle+24 \langle \epsilon \rangle+48 \beta _- \left(\beta _-^2+\beta _+^2\right)  \left(18 \beta _+-\delta _+\right)X_{\text{AB}}-48 \beta _- \delta _-(\beta_-^2+\beta_+^2)\right]\right.\nonumber\\
&\left.+\nu\left[\frac{123 \pi ^2}{256}-\frac{109}{2}+\bar{\gamma }_{\text{AB}} \left(\frac{49 \langle\bar{\beta }\rangle}{4}-\frac{3 \langle \delta \rangle}{16}-\frac{3 \beta _- X_{\text{AB}}}{4}+\frac{39 \beta _+}{4}+\frac{\delta _+}{2}-\frac{21 \pi ^2 \delta _+}{256}+\frac{15 \zeta }{4}-\epsilon _++\frac{225 \pi ^2}{512}-\frac{3353}{48}\right)\right.\right.\nonumber\\
&\left.\left.+\left(\frac{\langle\bar{\beta }\rangle}{4}+\frac{15 \pi ^2}{256}-\frac{2425}{96}\right) \bar{\gamma }_{\text{AB}}^2+\left(-\frac{71}{64}-\frac{21 \pi ^2}{1024}\right) \bar{\gamma }_{\text{AB}}^3+\left(\delta _++\frac{7}{4}-\frac{3}{4}\langle\bar{\beta}\rangle\right) \langle\bar{\beta }\rangle+ \left(\frac{9 \beta _- X_{\text{AB}}}{2}+\frac{81}{4}\right)\beta _+\right.\right.\nonumber\\
&\left.\left.+\frac{9}{4}( \beta _-^2+\beta _+^2)-\left(\frac{17}{12}+\frac{21 \pi ^2}{128}\right) \delta _+ +\frac{9 }{8}\langle \delta \rangle+ \left(-\frac{45 \beta _-}{4}-\frac{\delta _-}{8}+\frac{3 \epsilon _-}{8}\right)X_{\text{AB}}+\beta _- \delta _-+3 \zeta +\frac{\kappa _+}{2}-\frac{15 \epsilon _+}{8}+\frac{\langle \kappa \rangle}{4}-\frac{3}{8} \langle \epsilon \rangle\right.\right.\nonumber\\
&\left.\left.+\left(2\delta_++\frac{\bar{\gamma}_{\rm AB}(\bar{\gamma}_{\rm AB}+2}{2}\right)\ln\left(\frac{2}{p_{\infty}}\right)-\frac{3 \bar{\gamma }_{\text{AB}} \left(11 \left(\bar{\gamma }_{\text{AB}}+2\right){}^2-4 \langle \delta \rangle\right)}{32 \alpha _{\text{AB}} \left(\bar{\gamma }_{\text{AB}}+2\right)}+\frac{12 \left(\beta _-^2-\beta _+^2\right) \langle\bar{\beta }\rangle}{\bar{\gamma }_{\text{AB}}^2}\right.\right.\nonumber\\
&\left.\left.+\frac{\beta _+ \left(3 \langle \epsilon \rangle+4 \delta _- X_{\text{AB}}-8 \delta _+\right)}{2\bar{\gamma}_{\rm AB}}+\frac{\beta _- \left(X_{\text{AB}} \left(4 \delta _++3 \epsilon _+\right)-8 \delta _--3 \epsilon _-\right)}{2 \bar{\gamma }_{\text{AB}}}\right]\right.\nonumber\\
&\left.-\frac{3}{8}\nu^2\left[-\bar{\gamma }_{\text{AB}}^2-16 \bar{\gamma }_{\text{AB}}+3 -4 \beta _-^2+18 \beta _+-4 \delta _++6 \zeta -2 \epsilon _+-15\right]\right\}~,
\end{align}
\end{widetext}
where we use the notations of Refs.~\cite{Jain:2022nxs,Jain:2023fvt},
\begin{align}
&X_{AB}\equiv X_A-X_B \ , \\
&\langle\bar{\beta}\rangle\equiv -X_{AB}\beta_-+\beta_+,\\
&\langle\bar{\kappa}\rangle\equiv -X_{AB}\kappa_-+\kappa_+ \ , \\
&\langle{\delta}\rangle\equiv X_{AB}\delta_-+\delta_+~,\\
&\langle{\epsilon}\rangle\equiv -X_{AB}\epsilon_-+\epsilon_+ \ ,
\end{align}
with $X_{\rm A,B}\equiv m^0_{A,B}/M$. As the scattering angle is gauge-invariant, the arbitrary scale $\hat{s}$ has been cancelled between the two contributions as expected.

\section{Conclusions}
\label{Sec-conc}
Building upon the results of \cite{Jain:2022nxs,Jain:2023fvt,Julie:2022qux,Julie:2017pkb,Julie:2017ucp} for the corrections in the EOB metric coefficients $(A,B,Q_e)$ for massless scalar-tensor theory for the conservative part of the dynamics, we determined the scattering angle for hyperboliclike orbits upto the 3PN order for both the local-in-time and the nonlocal-in-time part of the dynamics. First, we compute the scattering angle for the local part of the dynamics by: (i) deriving the radial momentum as a function of $u$, orbital angular momentum and energy by iteratively solving the EOB energy conservation law; (ii) calculating the scattering angle using the standard techniques of Ref.~\cite{Damour:1988mr} for solving divergent integrals arising in the PN-expansion of the radial momentum except the integral $I_{\rm \chi}$ at the 3PN order; and (iii) computing the integral $I_{\rm \chi}$ by using the appropriate integration by parts and expanding in large-$j$ the remaining integral after change of variables \cite{Bini:2017wfr}.  We then computed the total contribution to the 3PN order scattering angle in the large-$j$ expansion.

Then, we computed the nonlocal-in-time contribution by using the approach introduced in Ref.~\cite{Damour:2015isa} for GR of order-reducing (time-localisation) the Hamiltonian in small-eccentricity case for hyperboliclike encounters \cite{Bini:2017wfr,Bini:2021jmj}.  Finally, we substituted the ST corrections of the metric potentials ($A,B,Q_e$) and sum both the local and nonlocal contributions in the large-$j$ expansion at 3PN order. As a test of our results, we checked that that the scattering angle coincides with the scattering angle of GR (see Ref.~\cite{Bini:2017wfr} for GR results) in the GR limit as expected.

This paper must be seen as a first step to compute the gauge-invariant scattering angle within the PN expansion for massless scalar-tensor theories. In future work we will address radiation reaction contributions to scattering. 

\begin{acknowledgements} 
The author is grateful to Piero Rettegno, Donato Bini and Thibault Damour for useful discussions and suggestions during the preparation of this work. The author also thank the hospitality and the stimulating environment of the Institut des Hautes Etudes Scientifiques.The author is jointly funded by the University of Cambridge Trust, Department of Applied Mathematics and Theoretical Physics (DAMTP), and Centre for Doctoral Training, University of Cambridge. The present research was also partly supported by the ``\textit{2021 Balzan Prize for 
Gravitation: Physical and Astrophysical Aspects}'', awarded to Thibault Damour. 
\end{acknowledgements}
\bibliography{refs.bib,local_st.bib}
\end{document}